\documentstyle[twoside,epsf,fleqn,espcrc2]{article}

\newcommand{\AmS}{{\protect\the\textfont2
  A\kern-.1667em\lower.5ex\hbox{M}\kern-.125emS}}

\newcommand{\beq}{\begin{equation}}
\newcommand{\eeq}{\end{equation}}        
\newcommand{\bqa}{\begin{eqnarray}}        
\newcommand{\eqa}{\end{eqnarray}}

\newcommand{\Expt}[1]{\langle #1 \rangle}

\newcommand{\Ha}{{\cal H}}
\newcommand{\Pa}{{\cal P}}

\title{Results From The UKQCD Parallel Tempering Project}

\author{B. Jo\'o\address{Department of Physics and Astronomy, 
                 University of Edinburgh, \\ 
                 James Clerk Maxwell Building, The Kings Buildings, 
                 Mayfield Road, Edinburgh EH9 3JZ, Scotland, UK},
        {\em UKQCD Collaboration}
}
\begin{document}

\begin{abstract}
We present results from our study of the Parallel Tempering algorithm.
We examine the swapping acceptance rate of a twin subensemble PT
system.  We use action matching technology in an attempt
to maximise the swap acceptance rate. We model the autocorrelation
times within Parallel Tempering ensembles in terms of autocorrelation
times from Hybrid Monte Carlo.  We present estimates for the
autocorrelation times of the plaquette operator.
\end{abstract}

\maketitle

\section{INTRODUCTION}
\label{s:Introduction}

Dynamical fermion simulations are still very demanding computationally. 
While a modern supercomputer is able to produce $O(100)$ quenched 
gauge configurations in an afternoon, months are needed to 
generate a similar number of dynamical configurations on the same machine.

Tempering algorithms have been successful in the past in
reducing autocorrelation times in difficult spin glass simulations
such as the random field Ising model \cite{MariPari}. 
It was of interest to see whether parallel tempering (PT) could achieve 
the same success for lattice QCD.

We present only our main results in this publication.
Full details of our work can be found in  \cite{PTpaper}. A lot
of the background information is covered in \cite{Match,BoydPT,Marinari}.

In section \ref{s:Algorithm} we outline the PT algorithm.
In section \ref{s:Autocorrelations} we present the predictions
of a simple model of autocorrelations in a twin subensemble PT system. Our
simulation parameters are outlined in \ref{s:Simulation}. Results from 
the simulations are presented in section \ref{s:Results}. Our summary and 
conclusions are in section \ref{s:Conclusions}

\section{THE ALGORITHM}
\label{s:Algorithm}

Parallel tempering consists of running several simulations, hereafter
referred to as {\em subensembles}, concurrently. 
Each subensemble has its own Hamiltonian $\Ha_i$,
parameter set and phase space.  The overall PT state then is the set
$\left\{ s_i | i = 1 ... N \right\}$ where $s_i$ is the state of
subensemble $i$ and $N$ is the number of subensembles. The PT phase
space is the direct product of the individual phase spaces.

The goal of PT is to construct a Markov Process which will converge to 
the equilibrium probability distribution
\begin{equation}
P_{eq}= \prod_i {1 \over Z_i} e^{-S_i},
\end{equation} 
where the terms on the right hand side
are the equilibrium distributions of the  individual subensembles
at their given parameter sets.

One defines two kinds of Markov transitions:
\begin{itemize}
\item
transitions within subensembles
\item 
transitions between subensembles
\end{itemize}

Transitions within subensembles are carried out using HMC. Transitions 
between subensembles involve a proposal to swap the current
fields in a subensemble $i$ with those in subensemble $j$. The swap proposal is
accepted with the Metropolis acceptance probability
\begin{equation} 
\Pa_s(i,j) = \mbox{min}\left( 1, e^{-\Delta \Ha} \right),
\end{equation} 
where
\begin{equation}
\Delta \Ha = \left\{ \Ha_j(a) + \Ha_i(b) \right\} 
            -\left\{ \Ha_i(a) + \Ha_j(b) \right\},
\end{equation}
which satisfies detailed balance with respect to $P_{eq}$ by
construction. The resulting overall Markov Process is connected and 
satisfies detailed balance with respect to the required equilibrium.

\subsection{Swap Acceptance}
The acceptance rate of swap proposals determines any reduction in
autocorrelation times over the usual HMC ones. It is also related to a
distance in parameter space via the formalism of action matching
technology \cite{Match}. It can be shown from detailed balance
considerations that the swap acceptance rate is given by
\begin{equation} 
\langle A \rangle = \mbox{erfc}\left( {1\over2}\sqrt{ \langle \Delta \Ha \rangle} \right). \label{e:AcceptRate}
\end{equation}

\section{AUTOCORRELATIONS}
\label{s:Autocorrelations}
Consider a twin subensemble system where both subensembles have the
same HMC autocorrelation function $C_H(t)$. This assumption is
justifiable if the HMC autocorrelation times vary slowly over the
region of parameter space where one intends to carry out PT
simulations. Table \ref{t:HMCAC} shows the integrated autocorrelation
times ($\tau_{\rm H}$) of the plaquette for two HMC simulations
separated by such a distance in parameter space. It can be seen that
the integrated autocorrelation times are equal within errors.

Let us assume that individual swap probabilities may be replaced by
the average swap probability $\Expt{A}$. Furthermore we are interested
only in an even number of swap attempts. Only after an even number of
successful swaps can a configuration end up in its original
subensemble.  We assume no cross correlations between subensembles.

It can be shown \cite{PTpaper} that the resulting
PT autocorrelation function in each subensemble is:
\begin{equation} 
C_{PT}(t) = \frac{1}{2} \left\{ 1 + \left( 1 - 2\Expt{A} \right)^{t} \right\} C_{H}(t)
\end{equation}
and assuming an exponentially decaying HMC autocorrelation 
function, the ratio of the HMC integrated autocorrelation time to its 
PT counterpart ($\tau_{\rm PT}$) is given by
\begin{equation}
\frac{\tau_{PT}}{\tau_{H}} = \frac{\Expt{A}(\tau_H-1)+1}{1+2\Expt{A}\tau_{H}}. 
\label{e:Autocorr}
\end{equation}
The above ratio is bounded above by $1$ and tends to $1 \over 2$ as
$\tau_H$ increases for a fixed $\Expt{A}$. Hence this model predicts
that the expected gain from PT over HMC is a factor of 2 in each
subensemble for roughly twice the work. Models of systems with
different HMC autocorrelation times are under investigation.
\begin{table}[t]
\setlength{\tabcolsep}{1.5pc}
\catcode`?=\active \def?{\kern\digitwidth}
\caption{HMC integrated autocorrelation times for the plaquette}
\label{t:HMCAC}
\begin{tabular*}{\columnwidth}{ccc}
\hline
 $\beta$ & $\kappa$ & $\tau_{\rm H}$ \\
 \hline 
$5.2$ & $.1335$ & $18(8)$ \\
$5.232$ & $.1335$ & $20(6)$ \\
\hline
\end{tabular*}
\end{table}
\section{SIMULATION}
\label{s:Simulation}
Our simulations used the GHMC \cite{ZbyshAndStephen} code developed by
the UKQCD collaboration for HMC transitions with extra logic to carry
out the swapping.  The simulation parameters to be tuned were the
inverse gauge coupling $\beta$, the fermion hopping parameter $\kappa$
and the clover coefficient $c$.

We carried out five twin subensemble simulations: $S1, S2, S3, S4$ and $S5$. The first subensembles of each one had parameters
$(\beta_1, c_1,\kappa_1) = (5.2, 2.0171, 0.1330)$.  The parameters for the
second subensembles are shown in table \ref{t:2PARAMS}. The simulation
parameters for $S1$, $S2$ and $S3$ were tuned using action matching
\cite{Match}. For $S4$ and $S5$ only $\kappa$ was varied. The lattice
size used was $8^3 \times 16$. A reference HMC run at the parameters
of the first ensembles was also carried out.
\begin{table}[b]
\setlength{\tabcolsep}{1.5pc}
\newlength{\digitwidth} \settowidth{\digitwidth}{\rm 0}
\catcode`?=\active \def?{\kern\digitwidth}
\caption{Simulation parameters for second ensembles}
\label{t:2PARAMS}
\begin{tabular*}{\columnwidth}{cc}
\hline
Simulation  & $(\beta_2, c_2, \kappa_2)$ \\
\hline 
$S1$      & $(5.2060, 2.01002, 0.13280)$ \\
$S2$      & $(5.2105, 2.00471, 0.13265)$ \\
$S3$     & $(5.2150, 1.99940, 0.13250)$ \\
\hline 
$S4$     & $(5.2, 2.0171, 0.13280)$ \\
$S5$     & $(5.2, 2.0171, 0.13265)$ \\
\hline
\end{tabular*}
\end{table}
\begin{table*}[t]
\setlength{\tabcolsep}{1.5pc}
\catcode`?=\active \def?{\kern\digitwidth}
\caption{Simulation Results}
\label{t:Results}
\begin{tabular*}{\textwidth}{@{}l@{\extracolsep{\fill}}cccccc}
\hline
\mbox{Simulation} & $\Delta \kappa (\times 10^{-4})$ & $\Expt{\Delta \Ha}$ & $\Expt{A}$ & $\tau_{\rm int}$ & $\frac{\tau_{\rm PT}}{\tau_{\rm H}}$ \\ \hline 
${\rm HMC}$ & - & -  & - & $26(6)$ & $1$ \\ \hline 
$S1$ & $-2.0$ & $1.23(2)$ & $0.43(1)$ & $12(3)$ & $0.5(2)$ \\ 
$S2$ & $-3.5$ & $3.76(4)$ & $0.17(1)$ & $19(4)$ & $0.7(2)$\\
$S3$ & $-5.0$ & $7.64(6)$ & $0.051(2)$ & $24(6)$ & $0.9(3)$ \\ 
$S4$ & $-2.0$ & $0.91(4)$ & $0.49(1)$ & $9(4)$ & $0.3(2)$ \\ 
$S5$ & $-3.5$ & $2.29(7)$ & $0.26(2)$ & $18(10)$ & $0.7(4)$ \\ \hline
\end{tabular*}
\end{table*}
\begin{figure}[h]
\epsfysize=65mm
\epsffile{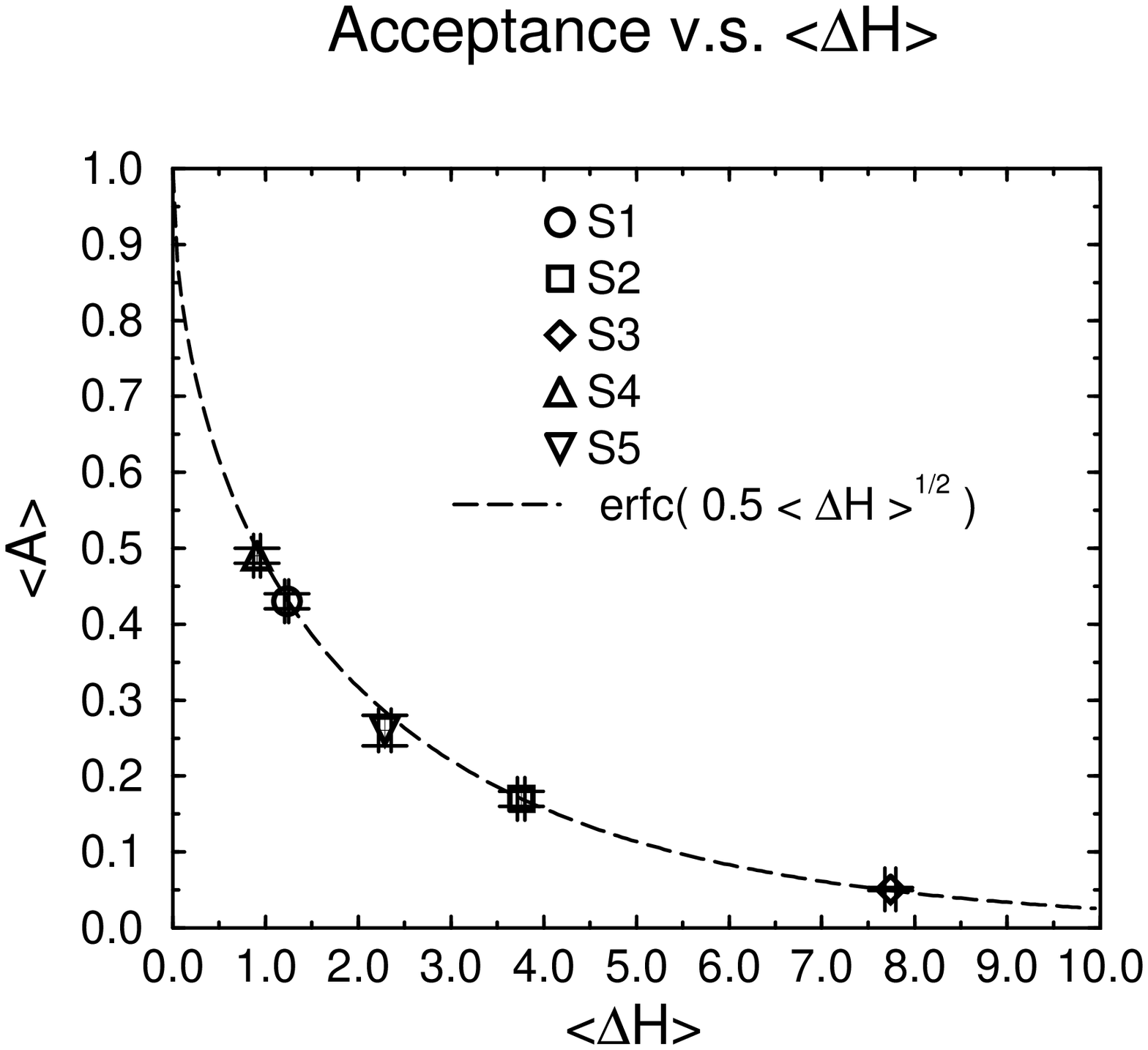}
\vspace{-0.8cm}
\caption{Swap Acceptance Rates}
\label{f:Accept}
\end{figure} 
\section{RESULTS}
\label{s:Results}
Table \ref{t:Results} summarises the results of our simulations.
Looking at columns 2, 3 and 4 it can be seen that $\Expt{\Delta \Ha}$
is usually greater than one and that $\Expt{A}$ drops rapidly as the
magnitude of $\Delta \kappa=\kappa_2-\kappa_1$ is increased. Figure
\ref{f:Accept} shows the acceptance rate $\Expt{A}$ as a function of 
$\Expt{\Delta \Ha}$. The dashed line is the graph of (\ref{e:AcceptRate}).
It can be seen that the measurements and the predictions of the 
acceptance rate agree very well. 

We note that the simulations with
parameters given by action matching technology ($S1 - S3$) have a
lower acceptance rate than the others. This is because the
fluctuations in $\Delta \Ha$ are larger using our pseudofermionic
Hamiltonian, than in the action used to perform the matching. This
issue is discussed more fully in
\cite{PTpaper}.

Column 5 in table \ref{t:Results} shows the integrated autocorrelation
times of the plaquette for our simulations.  Column 6 gives the
corresponding ratios of PT to HMC autocorrelation times. These ratios
are consistent with the predictions of the model described in section
\ref{s:Autocorrelations}.

\section{CONCLUSIONS}
\label{s:Conclusions}
We find that the acceptance rate of the PT algorithm drops very
rapidly with $\Delta \kappa$. This problem is expected to get worse on
larger lattices as $\Delta \Ha$, an extensive quantity, will have
larger fluctuations.  Over the range of available $\Delta \kappa$ the
HMC autocorrelation times are equal within errors and the predictions
of our autocorrelation model apply. We estimate that connecting fast
and slow decorrelating regions of parameter space would need a very 
large number of subensembles making PT impractical to use for lattice QCD with
currently available computer technology.
  
\section{ACKNOWLEDGEMENTS}
\label{Acknowledgements}
We  gratefully acknowledge funding PPARC under grant number GR/L22744 and 
EPSRC under grant number R/K41663.

\end{document}